\documentclass[aip,jcp,preprint]{revtex4-1}
\usepackage{graphicx}
\usepackage{amsmath}
\usepackage{amssymb}
\usepackage{xcolor}

\begin{document}

\title{Effect of Dielectric Discontinuity on a Spherical Polyelectrolyte Brush}

\author{Vinicius B. Tergolina}
\affiliation{Instituto de F\'isica, Universidade Federal do Rio Grande do Sul, Caixa Postal 15051, CEP 91501-970, Porto Alegre, RS, Brazil.}

\author{Alexandre P. dos Santos}
\email{alexandre.pereira@ufrgs.br}
\affiliation{Instituto de F\'isica, Universidade Federal do Rio Grande do Sul, Caixa Postal 15051, CEP 91501-970, Porto Alegre, RS, Brazil.}
\affiliation{Fachbereich Physik, Freie Universit\"at Berlin - 14195 Berlin, Germany.}

\begin{abstract}
In this paper we perform molecular dynamics simulations of a spherical polyelectrolyte brush and counterions in a salt free medium. The dielectric discontinuity on the grafted nanoparticle surface is taken into account by the method of image charges. Properties of the polyelectrolyte brush are obtained for different parameters, including valency of the counterions, radius of the nanoparticle and the brush total charge. The monovalent counterions density profiles are obtained and compared with a simple mean-field theoretical approach. The theory allows us to obtain osmotic properties of the system.
\end{abstract}

\maketitle

\section{Introduction}

The study of polyelectrolyte chains grafted to surfaces in a structure known as polyelectrolyte brush~(PEB) has acquired substantial interest recently, as covered by many reviews~\cite{Mi06,BaBo06,Ba07,JaBa09,BiMi12,SzTr14,DaBa15}. With the development of experiments with DNA molecules outside of the intracellular environment the study of cell-free gene expression has brought a new horizon for biotechnology. Examples go from double-stranded DNA brushes~\cite{KaTa14} to a single-step photolithographic biocompatible DNA mono-layer~\cite{BuDa08}, both on a biochip. In addition to these, its also valid to refer to other applications for the synthesization of PEBs such as protein absorption~\cite{SoPr98}, bioseparation~\cite{JiBa16} and targeted drug/gene delivery~\cite{WiAz07}. When referring to the term brush we assume that the grafting of the chains is dense enough in a way that the linear dimensions of the polyelectrolyte chains are much larger than the average distance between two neighboring charged polymers on the surface~\cite{Ba07}. Previous studies~\cite{Pi91,BoBi91,MeLa06,MeHo08} have shown that an essential property of a PEB is in its capability to confine a major quantity of counterions in a way to compensate its electrical charge, resulting in high osmotic pressure governing its stretching dynamics. Besides their extensive range of applications, PEBs have been studied in a range of different configurations as well, they can be generated either by grafting polyelectrolyte chains to planar~\cite{AhFo97,BiRu99,CuSi99,RuBa04,BePo04} or to strongly curved systems as e.g. cylinders or spheres~\cite{ZhYu96,BiHa97,GuMu98,HaBi98,GuWe99,MuFo01}, the last being the focus of our present work. Recently, different types of neutral polymer brushes with interesting properties composed by dipolar ions called zwitterions~\cite{ChBr11,KoTe13} in the form of polyzwitterions, have attracted attention and deserve citation.

In relation to spherical PEB we can elicit their main structure as being formed by an inorganic core nanoparticle and an organic layer/shell in the form of polyelectrolyte chains grafted to its surface. As a result of their mechanical stability, high surface area, and ease of synthesis, silica/polymer hybrid nanoparticles have been studied more extensively~\cite{RaPa12,JiBa16}. Spherical PEBs also carry a number of advantages in comparison with planar ones. They can be studied by a wide variety of distinct methods coined for colloidal particles investigation, from scattering methods~\cite{GrLa00,GuBa01,DiPa04,ScPa09} to, more recently, dielectric spectroscopy~\cite{GuZh16}. Furthermore, the colloidal dimensions of the spherical PEB may be used to create well defined surfaces of the order of many $m^{2}$ that can be used for nanoparticle/protein immobilization~\cite{WiHa03,ShBa04}, and they can also be viewed as models for the study of carboxylated latex particles that constitute a major industrial product~\cite{Di99}.

Amongst previous works, we can cite efforts to theoretically describe PEBs~\cite{RaGo14,ErDe16,NaNe03}. Regarding molecular dynamics~(MD) simulations of spherical PEBs we can cite, as few examples, studies on the dependence of the brush thickness due to different parameters and conformations~\cite{CsSe00,Se03,SaCa07}, studies on brush size as a function of chain lengths, salt concentrations~\cite{KuSe05} and grafting densities - these accompanied by comparisons to mean field or self-consistent field theories~\cite{HeMe10,MeLi15}. The effect of multivalent ions on brush conformations was also extensively studied~\cite{MeLa06,TiMa09,YaXu09,LiPi17}. Nevertheless, MD simulations of PEBs that take into account the dielectric discontinuity between the grafted nanoparticle and surrounding medium are unprecedented so far, to the best of our knowledge. In spite of the preceeding statement, the problem of charged particles in heterogeneous dielectric media has been broadly studied resulting in the coinage of different methods. Among those we refer to threatments which can be applied in the spherical geometry for applications in colloidal science. Even if the computational cost is high, one can use Legendre polynomials~\cite{Me02,GuDe14} to perform MC or MD simulations. A variational formulation has gain attention lately as a more general method for the solution of the Poisson equation treating the local polarization charge density as a dynamic variable~\cite{JaCr12,JaCr13}. A more efficient method considers the images and uniformly distributed counter-image charges inside the dielectric void as an approximation which works very well for low dielectric constants~\cite{DoBa11,BaDo11}. In this work we intend to include nanoparticle polarization using the previously mentioned method and perform MD simulations of a PEB in a salt free suspension. In addition, a simple Poisson-Boltzmann~(PB) theory is presented in order to account for the counterions concentration in mean-field regime.

In the next section we explain the model and simulation method followed by the presentation of the theory developed for weak electrostatic coupling. The results are presented in the further section. In the last section we finish describing the conclusions of the present work and general perspectives.

\section{Model and Simulation Method}

We follow a standard coarse grained model for the polymer chains and counterions confined in a spherical cell of radius $R$. The $N_p=14$ chains are represented by $N_m$ charged hard spheres (monomers) of radii $r_m = 2~$\AA\ and charge $+q$, where $q$ is the proton charge. The first monomers of the chains are grafted to the surface of a sphere of radius $a$ and relative dielectric constant $\epsilon _{c}$, representing the nanoparticle base particle. The first monomers are all uniformly distributed on the nanoparticle surface, grafted at distance $r_m$ from it. The $N_c$ counterions are modeled as hard spheres with effective radii $r_c=r_m$ and charge $-\alpha q$, where $\alpha$ is the valency. The number of counterions is defined as $N_c=N_p N_m/\alpha$ in order to keep the system with zero total charge. The medium in which the polyelectrolyte is immersed is represented by structureless water with relative dielectric constant $\epsilon_{w}=80$. The Bjerrum length, defined as $\lambda_{B} = q^{2}/\epsilon _{w}k_{b}T$, is $7.2~$\AA, the value for water at room temperature. 

Following a method previously developed~\cite{DoBa11}, we investigate the influence of the nanoparticle polarization by means of image charges. The calculation of image charges for the spherical geometry is not as straightforward as for the planar geometry. The continuity of the tangential component of the electric field and of the normal component of the displacement field across the nanoparticle-water interface, requirements of the Maxwell equations boundary conditions, give rise to a counter-image line charge in addition to the punctual image charge that is the usual requirement for planar geometry~\cite{No95}. The electrostatic potential at an arbitrary position ${\bf{r}}$ produced by the arbitrary charge $q_i$ located at ${\bf r}_i$ outside the nanoparticle is approximated by
\begin{eqnarray}
\phi({\bf{r}};{\bf r}_i) = \frac{q_i}{\epsilon_w|{\bf r}-{\bf r}_i|} + \frac{\gamma q_i a}{\epsilon_w r_i|{\bf{r}}-\frac{a^{2}}{{r^{2}_{i}}}{\bf r}_i|} + \nonumber \\
\frac{\gamma q_i}{\epsilon_w a} \log\left(\frac{r r_{i}-{\bf{r}}\cdot{\bf r}_i}{a^2 - {\bf{r}}\cdot{\bf r}_i+\sqrt{a^4-2a^{2}({\bf{r}}\cdot{\bf r}_i)+r^{2}r^{2}_{i}}}\right) \ ,
\end{eqnarray}
where $r_i=|{\bf r}_i|$, $r=|{\bf r}|$ and $\gamma = (\epsilon_{w} - \epsilon_{c})/(\epsilon_{w}+\epsilon_{c})$. This expression is valid~\cite{DoBa11} for $\epsilon_{w} >> \epsilon_{c}$ . The total electrostatic energy is
\begin{eqnarray}
U_{elec} = \sum_{i=1}^{N-1} \sum_{j=i+1 }^{N} q_j \phi({\bf r}_j;{\bf r}_i) + \nonumber \\
\sum_{i=1}^N \frac{\gamma q_i^2 a }{2 \epsilon_w (r_i^2-a^2)} + \sum_{i=1}^N \frac{\gamma q_i^2 \log{(1-\frac{a^2}{r_i^2})}}{2 \epsilon_w a} \ .
\end{eqnarray}
The two last terms above are the ionic electrostatic self energy.

The elastic bonds between adjacent monomers of the same chain in the brush are modeled by the following non linear energy potential~\cite{DiPa05,QuMa13,LuMa14,SaLe16},
\begin{equation}
U_{bond} = \sum_{ad. mon.} \dfrac{A}{2}(r-r_{0})^{2} \ ,
\end{equation}
where $r=|{\bf r}_i-{\bf r}_j|$ is the distance between adjacent monomers $i$ and $j$. The sum is made over all adjacent monomers of the same polymer chains, $A=0.9k_{B}T$ and $r_{0}=5~$\AA, following the aforementioned reference~\cite{SaLe16}.

The total force acting on the charged specie $k$ is
\begin{equation}
{\bf F}_k = -\nabla_{{\bf r}_k} (U_{elec}+U_{bond}) \ .
\end{equation}
%%%%%%%%%%%%%%%% figure %%%%%%%%%%%%%%%%%%%%%
\begin{figure}[t]
\begin{center}
\includegraphics[scale=0.45]{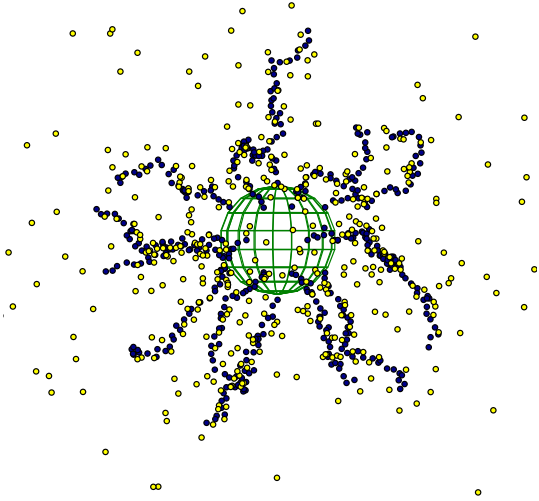}%\vspace{0.1cm}
\end{center}
\caption{Representation of the spherical PEB. Darker spheres represent monomers, while lighter spheres represent counterions.}
\label{fig0}
\end{figure}
%%%%%%%%%%%%% end of figure %%%%%%%%%%%%%%%%%
The molecular dynamics simulations were performed for constant time steps by means of well known Langevin equation~\cite{AlTi87},
\begin{equation}
{\bf p}'_i(t) = {\bf F}_i(t) - \Gamma {\bf p}_i(t)+{\bf R}_i(t) \ ,
\end{equation}
where ${\bf p}_i(t)$ is the momentum of particle $i$ at time $t$, ${\bf F}_i(t)$ is the force felt by this particle, $\Gamma$ is the friction coefficient and ${\bf R}_i(t)$ is the stochastic force acting on particle $i$ which satisfy the fluctuation dissipation relation. The Verlet-like method developed by Ermak~\cite{AlTi87} is used to solve previous equation.

The mechanism chosen to avoid the superposition between all particles and surfaces is a hard sphere potential. This was preferred over a caped Lennard-Jones type potential for the reason that the latter was tested showing little to no difference from the hard spheres potential while requiring time steps much smaller to advert simulation crashes. In Fig.~\ref{fig0}, a snapshot of MD simulations after equilibrium is shown for monovalent counterions.

\section{Theory}

At room temperature, electrostatic correlations between monovalent ions can be neglected~\cite{Le02}. A mean-field PB equation is used to obtain the density profile of counterions. We do not consider the dielectric discontinuity on the nanoparticle-water interface because this effect is very small in this regime, as it will be shown in the Results section. Also, the qualitative results obtained with the present method allows us to consider this approximation. However, it is important to mention that this effect can be important when more accuracy is necessary in the study of the electric double layer~\cite{DoBa11,BaDo11}.

% The solution of linear PB equation was obtained for proper boundary conditions in planar geometry; using charging process and some curvature corrections the potential energy was idealized~\cite{BaDo11}. The energy required to bring an ion from infinity to a distance $z$ from nanoparticle surface is given by~\cite{BaDo11}
% \begin{eqnarray}
% \beta W(z)=[\beta W_0 \frac{r_c}{z} - \frac{\lambda_B}{4 z} + \frac{a \lambda_B}{2 (z^2+2az)}+\nonumber \\
% \frac{\lambda_B}{2 a} \log{(1-\frac{a^2}{(a+z)^2}})]e^{-2 \kappa(z-r_c)} \ ,
% \end{eqnarray}
%%%%%%%%%%%%%%%% end of equation %%%%%%%%%%%%%%%%%%%%%
% where
%%%%%%%%%%%%%%%% equation %%%%%%%%%%%%%%%%%%%%%
% \begin{equation}
% \beta W_0=\frac{\lambda_B }{2}\int_0^\infty dk\frac {k[s\ \mathrm{cosh}(k r_c)-k\ \mathrm{sinh}(k r_c)]}{s[s\ \mathrm{cosh}(k r_c)+k\ \mathrm{sinh}(k r_c)]} \ ,
% \end{equation}
%%%%%%%%%%%%%%%% end of equation %%%%%%%%%%%%%%%%%%%%%
% $s=\sqrt{k^2+\kappa^2}$, $\kappa=\sqrt{\dfrac{4 \pi \lambda_B N_c}{ V}}$, $V$ is the volume accessible to counterions and $z=r-a$. In results section we show that the nanoparticle polarization is not very important in the case of monovalent ions at room temperature. However, we decided to include the image potential in the theory as it corrects some small deviations near surface.

%%%%%%%%%%%%%%%% figure %%%%%%%%%%%%%%%%%%%%%
\begin{figure}[t]
\begin{center}
\includegraphics[scale=0.3]{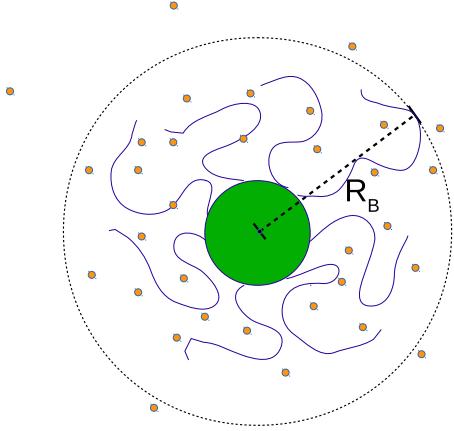}%\vspace{0.1cm}
\end{center}
\caption{Definition of effective PEB radius, $R_B$.}
\label{fig1}
\end{figure}
%%%%%%%%%%%%% end of figure %%%%%%%%%%%%%%%%%
The charge distribution of PEB is constructed as if all the monomers are aligned with the nanoparticle center, with effective distance between them equals to $r_{ef}=0.75 r_{m}$ to account for the bending of the chains. The modified PB equation takes the form
\begin{eqnarray}\label{mpb}
\nabla^2 \phi(r)=-\frac{4\pi}{\epsilon_w}\left[ \sum_{i=1}^{N_m}\sigma_i\delta(r-r_i) - q \alpha  \rho(r) \right] \ ,
\end{eqnarray}
where $\phi(r)$ is the mean electrostatic potential, $\sigma_i$ are the charge densities of the corresponding layers of monomers, given by $\sigma_i=N_p/4\pi r_i^2$ where $r_i=a+r_{m}+(i-1) 2 r_{ef}$.
%%%%%%%%%%%%%%%% equation %%%%%%%%%%%%%%%%%%%%%
The counterions density profile is given by
\begin{eqnarray}
\rho(r)=N_c\frac{e^{-\beta \alpha q \phi(r)}}{4\pi\int_{(a+r_c)}^R dr' r'^2 e^{-\beta \alpha q \phi(r')}} \ .
\end{eqnarray}
% At this point it deserves to mention that the Eq.~\ref{mpb} can be calculated setting $W(z)=0$, as the image effects are not very important for the presented parameters. The main property to determine counterions distributions is the number of monomers in chains.

The solution of Eq.~\ref{mpb} is performed by Picard iterative process.

\section{Results}

The results are presented in the form of average particles concentration profiles and average effective PEB radius (the distance between the center of the grafted nanoparticle and the more distant monomer), see Fig.~\ref{fig1}.
%%%%%%%%%%%%%%%% figure %%%%%%%%%%%%%%%%%%%%%
\begin{figure}[t]
\begin{center}
\includegraphics[scale=0.25]{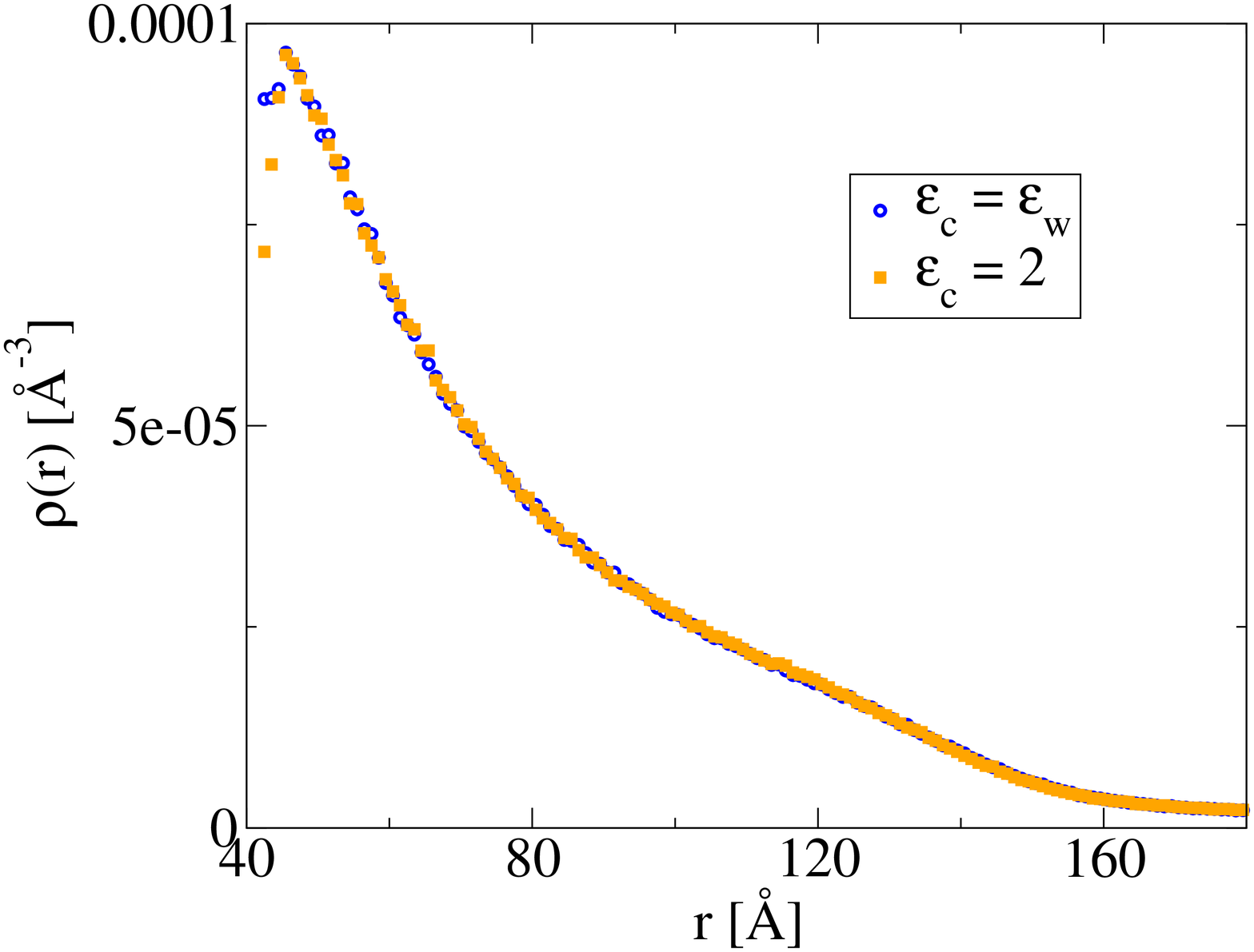}%\vspace{0.1cm}

\includegraphics[scale=0.25]{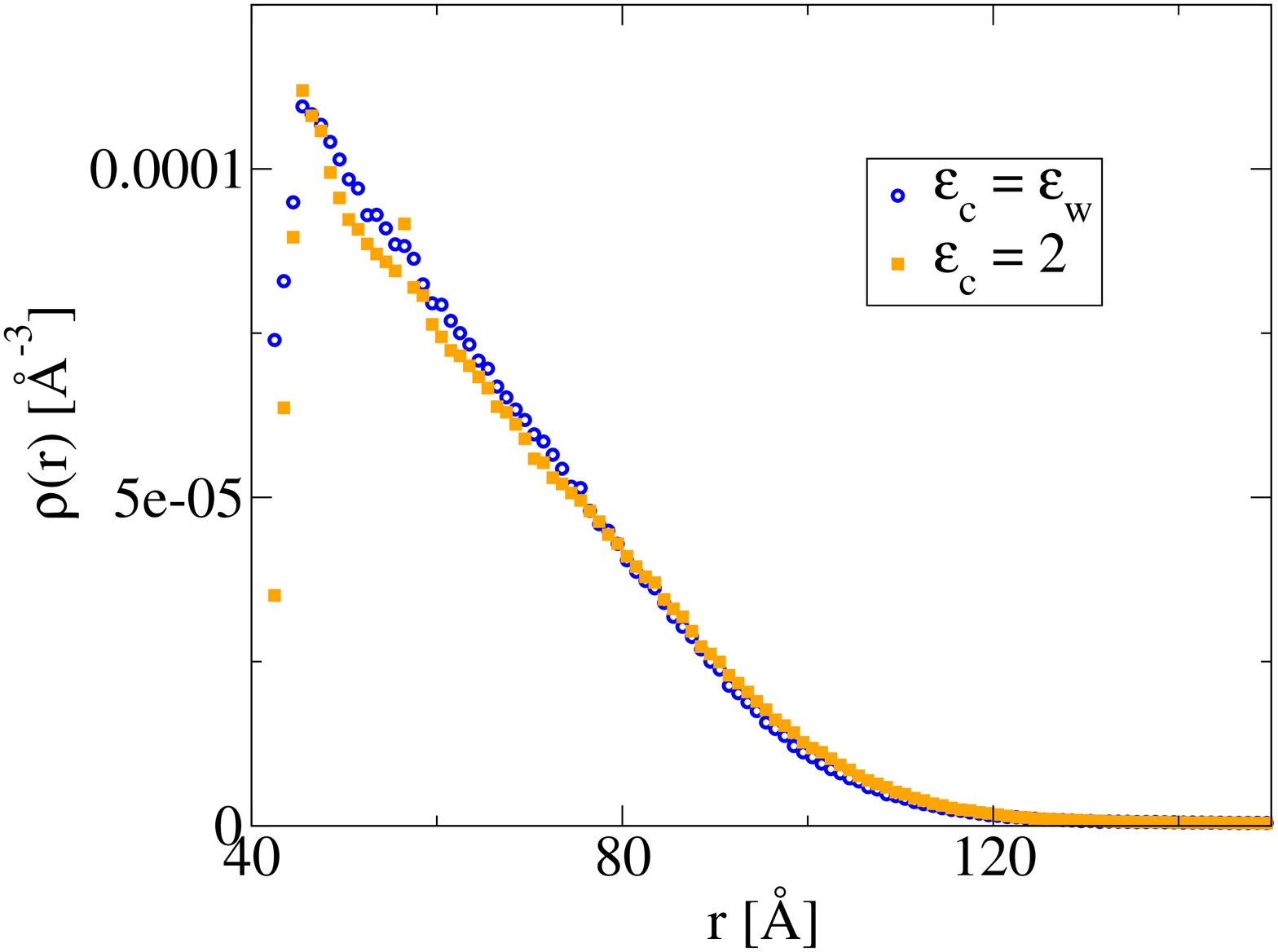}%\vspace{0.1cm}

\includegraphics[scale=0.25]{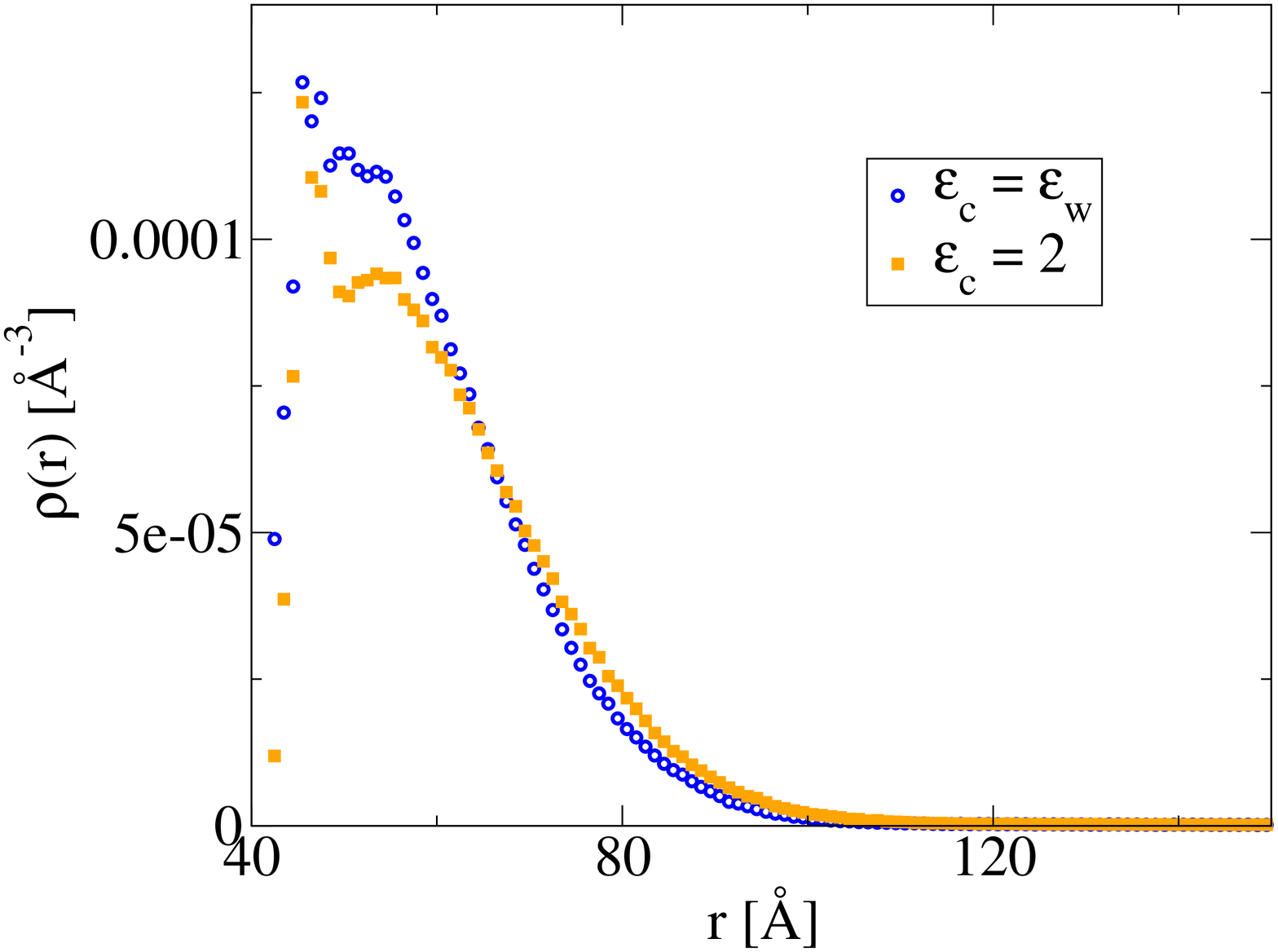}%\vspace{0.1cm}
\end{center}
\caption{Density profiles of
counterions obtained for $\alpha = 1,2,3$, from top to bottom, respectively. Polyelectrolyte brush individual chains with $N_{m} = 30$ and nanoparticle radius $a = 40~$\AA.}
\label{fig2}
\end{figure}
%%%%%%%%%%%%% end of figure %%%%%%%%%%%%%%%%%
We start by studying the effect of the dielectric discontinuity on the counterion distribution around the PEB, see Fig.~\ref{fig2}. We choose the following values for the nanoparticle relative dielectric constant, $\epsilon_{c} = 2$ and $\epsilon_{c} = \epsilon_{w}$. Whereas on the first choice we choose the typical dielectric constant value of silica, on the second case we ignore the dielectric discontinuity by having the nanoparticle represented by the same material as the medium in which it is inserted, water. Silica nanospheres coated with polymer brushes have already been used for effective separation of glycoproteins~\cite{JiBa16}.

%%%%%%%%%%%%%%%% figure %%%%%%%%%%%%%%%%%%%%%
\begin{figure}[t]
\begin{center}
\includegraphics[scale=0.3]{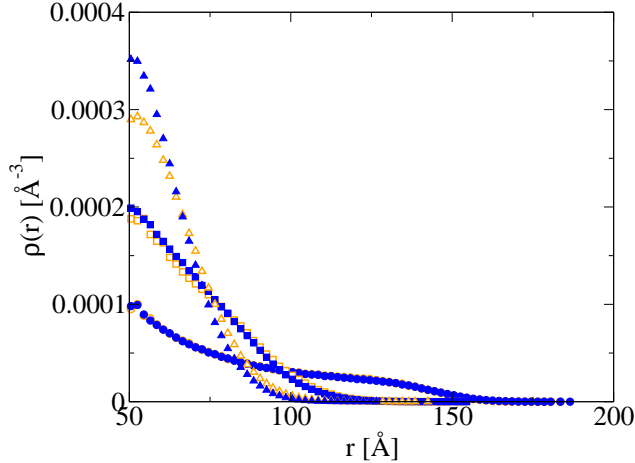}\vspace{0.2cm}
\end{center}
\caption{Density profiles of monomers obtained for $\alpha = 1,2,3$, circles, squares and triangles, respectively. Full symbols represent $\epsilon_c=\epsilon_w$, while open symbols $\epsilon_c=2$. The parameters are the same as in Fig.~\ref{fig2}.}
\label{fig3}
\end{figure}
%%%%%%%%%%%%% end of figure %%%%%%%%%%%%%%%%%
The influence of the dielectric discontinuity on monovalent ions is very small and most of the pattern we see is caused by osmotic pressure inside the brush, which tends to repel counterions. Similar brush configuration have been extensively explored before without consideration for dielectric discontinuity so that the results showing that multivalent counterions are more deeply absorbed are expected. Although the density maximum concur for both distributions of multivalent ions, we find that the polarization of the silica nanoparticle tends to broaden their distributions since they feel more repelled by their image charges. Also, the effective brush radius, $R_B$, tends to be higher due to image charges of chains, which can affect the ionic distribution far away from brush. For charged nanoparticles and surfaces the consequence of a dielectric discontinuity in ionic distribution is very local, near surfaces~\cite{DoBa11,BaDo11,JaCr13,DoGi16}.
% It is also interesting to notice the double maxima pattern that rises subtly for the bivalent, and in a more pronounced way, for the trivalent ions. We can see that the inclusion of the calculation of image charge potentials enhances deeply this pattern for the case of the trivalent ions giving two well defined prefered positions for the counterions inside the brush.
The importance of the polarization effect for the trivalent case can also be observed in the density profiles of monomers, see Fig.~\ref{fig3}.

\begin{figure}[t]
\begin{center}
\includegraphics[scale=0.3]{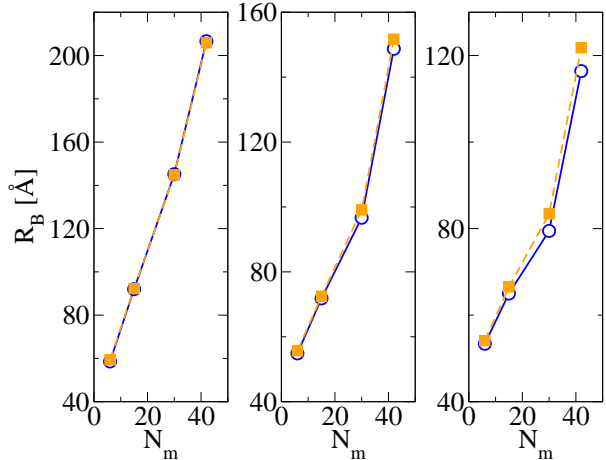}%\vspace{0.1cm}
\end{center}
\caption{PEB average effective radius as function of $N_m$ for $a = 40~$\AA\ and $\alpha = 1,2,3$, from left to right panels, respectively. The circles represent the case which $\epsilon_c=\epsilon_w$, while squares, $\epsilon_c=2$.}
\label{fig4}
\end{figure}

Moving further, we study the brush behavior over different number of monomers and different counterion valence by calculating $R_B$, see Fig.~\ref{fig4}. We define the PEB radius, $R_B$, as the average distance between the center of grafted nanoparticle and the more distant monomer.
%%%%%%%%%%%%%%%% figure %%%%%%%%%%%%%%%%%%%%%
%%%%%%%%%%%%% end of figure %%%%%%%%%%%%%%%%%
Here we confirm that image charges have little to no influence over the brush diameter for monovalent counterions. This is not the case for larger brushes composed by $30$ and, more explicitly, $42$ monomers, surrounded by multivalent ions. In this case we can find a considerable increase in $R_B$ when accounting for the dielectric discontinuity when compared to the homogeneous case. The polyelectrolyte chain total charge is high for a sufficient number of monomers and they are, by construction, near the nanoparticle surface. This means that image charges play important role in the brush radius value when this value is sufficiently small. The difference in both approximations can achieve $\approx 9\%$ for the discussed parameters. The smaller values obtained for $R_B$ in the case of multivalent ions are in agreement with experiments which relate the collapse of the spherical PEB with the addition of multivalent ions in solution~\cite{MeLa06,PlWa07}.

%%%%%%%%%%%%%%%% figure %%%%%%%%%%%%%%%%%%%%%
\begin{figure}[t]
\begin{center}
\includegraphics[scale=0.3]{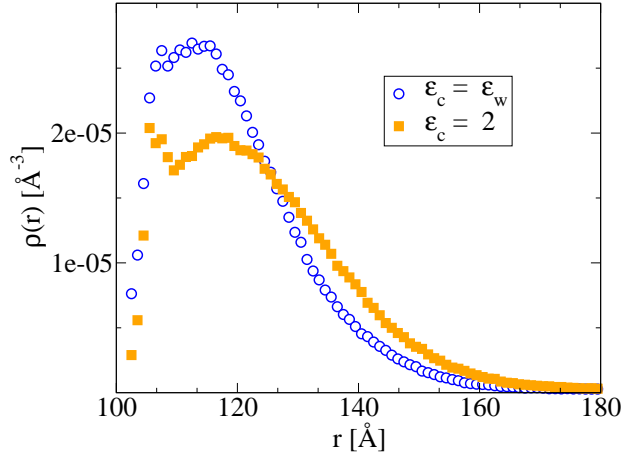} %\vspace{0.1cm}
\end{center}
\caption{Density profiles of counterions obtained for $\alpha = 3$ for two relative dielectric constants of the nanoparticle. Polyelectrolyte brush individual chains with $N_{m}=30$ and grafted nanoparticle radius $a=100~$\AA.}
\label{fig5}
\end{figure}
%%%%%%%%%%%%% end of figure %%%%%%%%%%%%%%%%%
The density profile for a special case in which the brush's nanoparticle is relatively big ($a = 100~$\AA) is shown in Fig.~\ref{fig5}. The polarization of the nanoparticle undoubtedly has a strong influence over the trivalent counterions profile, showing that the role played by nanoparticle polarization is not only to further fend the colloid and the counterions but also to spread their distribution, in comparison with the unpolarized nanoparticle. Its also worthy to remark the double peak pattern present in the $\epsilon_{c} = 2$ curve, much more protruding than in the $\epsilon_{c} = \epsilon_{w}$ curve, indicating two clear preferred regions for the trivalent counterions. This is a competition between the electrostatic interaction of multivalent ions with the entire brush and with their local chain, see also Fig.~\ref{fig2}, bottom panel. The polarization of the nanoparticle separates more explicitly these regions as a result of the shifting of the ionic distribution.

In order to measure the effect of nanoparticle polarization on counterions distributions as a function of nanoparticle curvature we calculate the relative difference between profiles defined as $\Delta=\dfrac{\sqrt{\int_a^R dr [\rho_2(r)-\rho_{\epsilon_w}(r)]^2}}{\int_a^R dr \rho_{\epsilon_w}(r)}$, where $\rho_{\epsilon_w}(r)$ is the counterion profile for $\epsilon_c=\epsilon_w$ and $\rho_2(r)$ for $\epsilon_c=2$. We take the cases of Fig.~\ref{fig2} and similar ones except for the parameters $a=80~$\AA\ and $R=500~$\AA\ for comparison. We set these lengths in order to maintain constant volume fraction in the comparison. The volume fraction is defined as $\phi_{frac}=a^3/R^3$. For $\alpha=1$ we obtain for $\Delta$ the values $0.0041$ and $0.0032$. For divalent $\alpha=2$ sets we obtain $0.0115$ and $0.0166$. The values found for $\alpha=3$ were $0.0265$ and $0.0252$, all numbers for $a=40~$\AA\ and $a=80~$\AA, respectively. These results show us that there is no influence of the nanoparticle curvature in the polarization effect on the counterions distribution for constant PEBs volume fraction. However, if we take for comparison two sets with the same cell radius $R$ but with different nanoparticles radius $a$, the curvature can decrease the effect of dielectric discontinuity on the counterions distribution. We take the trivalent case of Fig.~\ref{fig2} and the set of Fig.~\ref{fig5}. The parameters are the same with the exception of the nanoparticle radius which is $40~$\AA\ and $100~$\AA, respectively. We then obtain the values $0.0265$ and $0.0424$, respectively, for the profiles relative difference, showing that the decreasing in curvature enhance the aforementioned effect.
%%%%%%%%%%%%%%%% figure %%%%%%%%%%%%%%%%%%%%%
\begin{figure}[t]
\begin{center}
\includegraphics[scale=0.25]{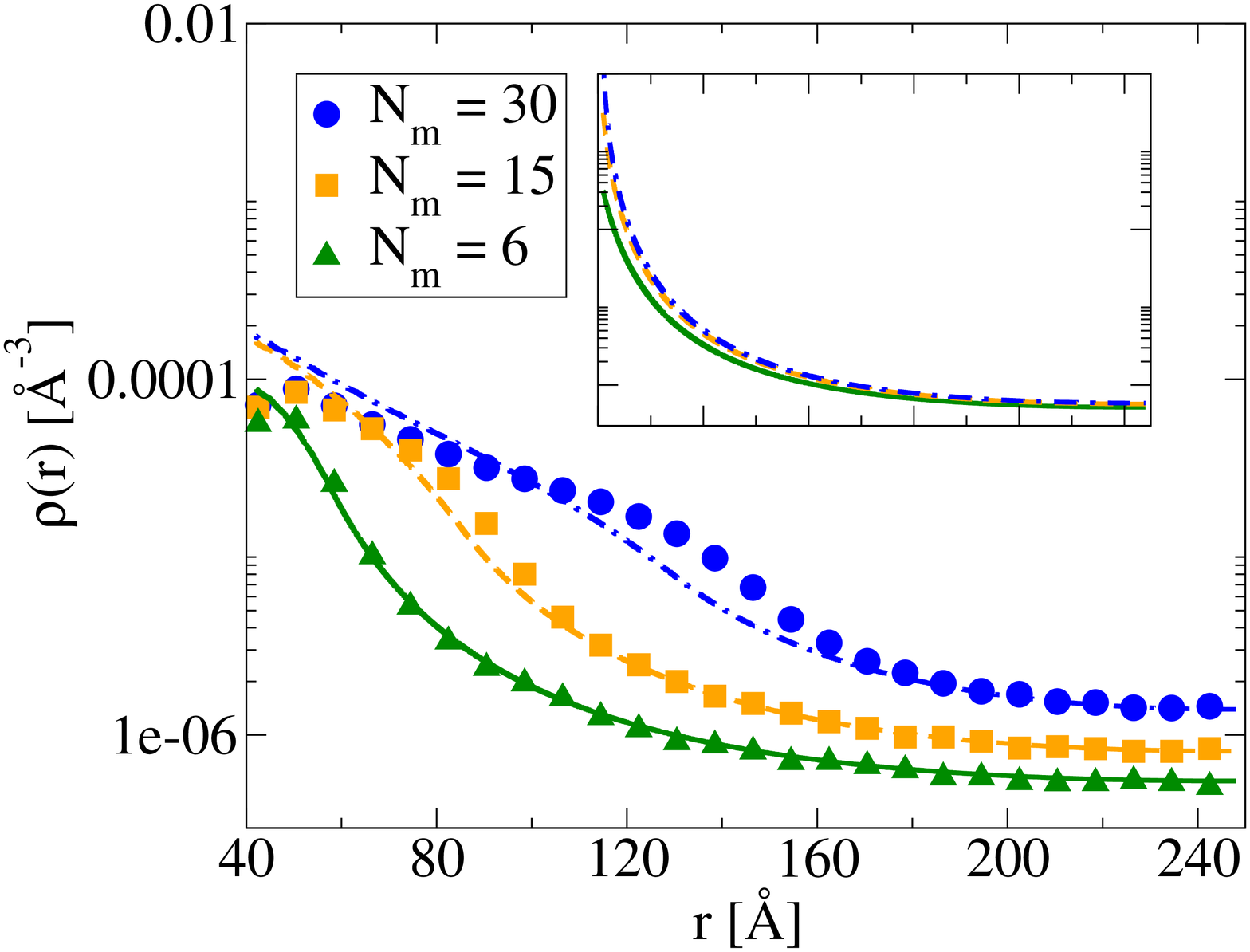} %\vspace{0.1cm}
\end{center}
\caption{Density profiles of counterions obtained for $\alpha = 1$, $a=40~$\AA\ and various number of monomers, $N_m$. The lines represent the results of the present theory, while symbols the results of simulations. The inset shows the solutions of PB equation if all the charged monomers are located on the nanoparticle surface, for the same parameters and the same $x$ and $y$ axis scales.}
\label{fig6}
%\vspace{0.5cm}
\begin{center}
\includegraphics[scale=0.25]{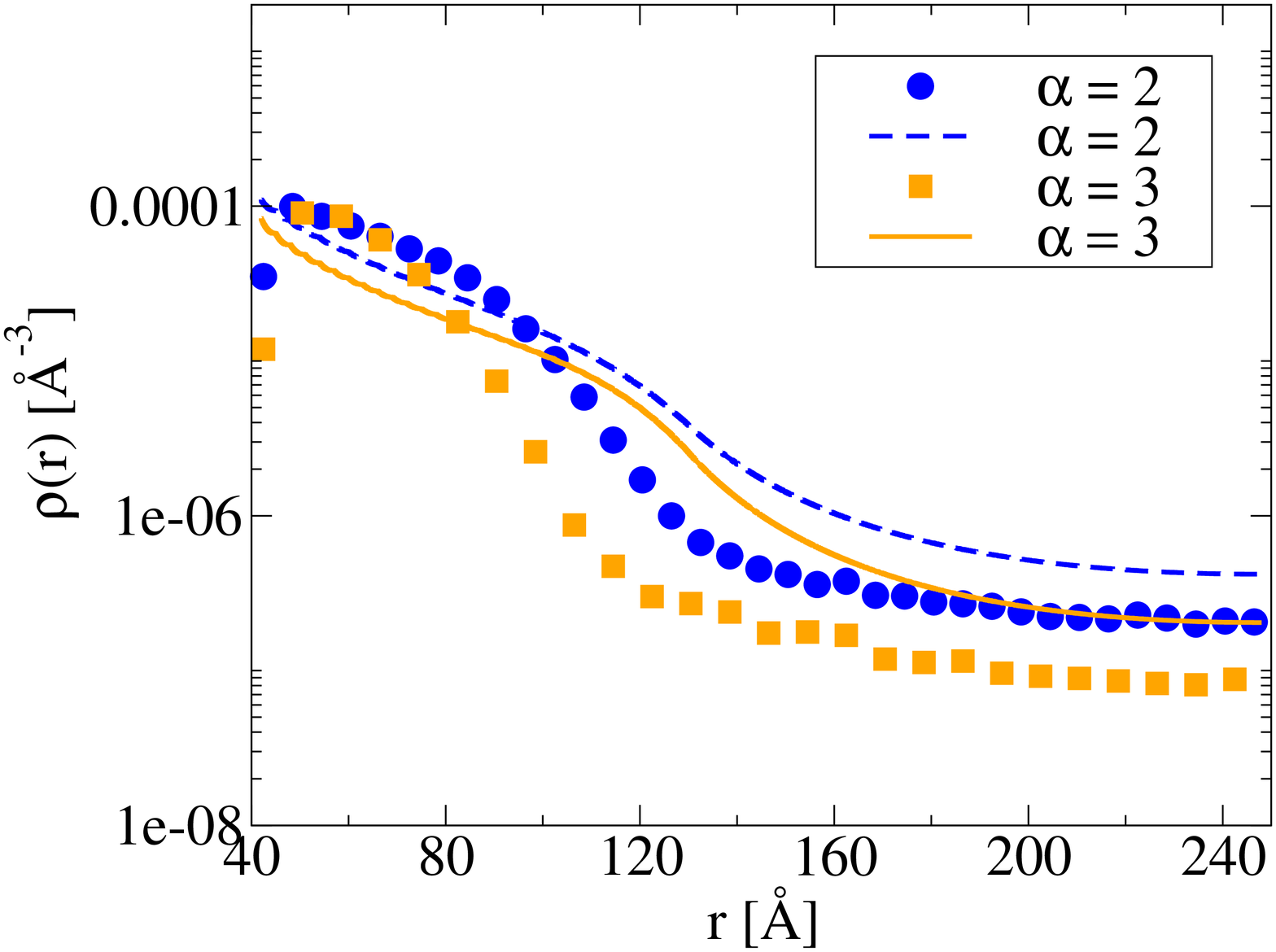} %\vspace{0.1cm}
\end{center}
\caption{Density profiles of multivalent counterions obtained for $N_m = 30$ and $a=40~$\AA. The lines represent the results of the present theory, while symbols the results of simulations.}
\label{fig7}
\end{figure}
%%%%%%%%%%%%% end of figure %%%%%%%%%%%%%%%%%

% Considering that the free space between polyelectrolytes on the brush plays a main role in the absortion of counterions, and therefore in the dynamics of the polymer chains, we proced study different simulations with different colloid radius as a way to increase or decrease individual chain's distance. 
%%%%%%%%%%%%%%%% figure %%%%%%%%%%%%%%%%%%%%%
% \begin{figure}[h]
% \begin{center}
% \includegraphics[scale=0.3]{lenXRcol30C3.eps}%\vspace{0.1cm}
% \end{center}
% \caption{Comparisson between the PEB radius and different colloid radius $a$ for $N_{m} = 30$, $\alpha = 3$.}
% \label{fig4}
% \end{figure}
%%%%%%%%%%%%% end of figure %%%%%%%%%%%%%%%%%

% The conformational states avaliable for the brush are clearly altered by the variation of the colloid size, see Fig.~\ref{fig4}. With the increase of $a$, the charge density of the brush decreases as the number of chains and monomers is fixed, what brings less counterions to chains, decreasing the electrostatic shielding and allowing an increase in brush radius. As we can observe ...

We move further in the results section by comparing simulations with the present theory for monovalent ions, see Fig.~\ref{fig6}. The theory is not able to describe properly the monovalent counterions structure around the brush, except for shorter chains. However, the agreement is very good in the region far from nanoparticle surface, for the studied chain lengths. The present method allows us to quantitatively account the adsorption of monovalent counterions, which means that osmotic properties of a brush suspension can be studied using the present method. We can define, for example, effective charges of PEBs, subject for a future work. It is important to mention the interesting effect that the boundary ionic concentrations are not saturated with the increase in the macroparticle charge as it is observed in colloidal suspensions, see inset of Fig~\ref{fig6}. This saturation observed in colloidal suspensions reflects the independence of the colloidal effective charge with the colloidal charge~\cite{AlCh84,TrBo03}. This is not the case for PEBs as can be seen in Fig.~\ref{fig6}. For multivalent counterions, as expected, the theory is not able to describe the asymptotic curve, as can be seen in Fig.~\ref{fig7}, not even by reasonably decreasing the value of $r_{ef}$. The counterion-counterion and counterion-monomer electrostatic correlations take place and the present mean-field theory is not able to account for these effects.

%%%%%%%%%%%%%%%% figure %%%%%%%%%%%%%%%%%%%%%
\begin{figure}[t]
\begin{center}
\includegraphics[scale=0.3]{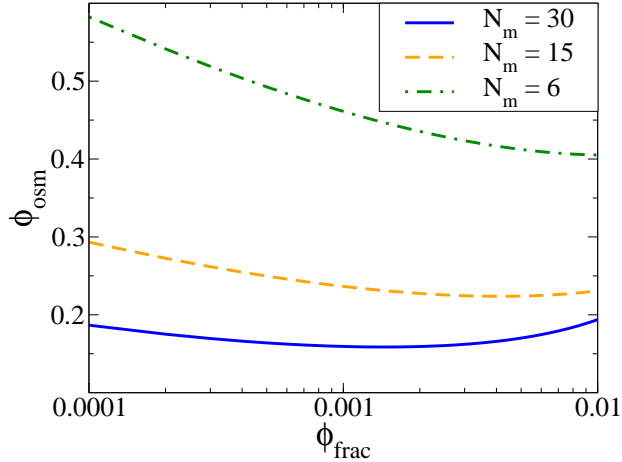} %\vspace{0.1cm}
\end{center}
\caption{Osmotic coefficient versus volume fraction for the same parameters of Fig.~\ref{fig6}.}
\label{fig8}
\end{figure}
%%%%%%%%%%%%% end of figure %%%%%%%%%%%%%%%%%
As an application of the method for monovalent counterions, we calculate the osmotic coefficient which is defined as the fraction between the pressure and ideal pressure given by $\phi_{osm}=\rho_{bulk}/\rho_{id}$, where $\rho_{bulk}$ is the counterion bulk concentration and $\rho_{id}=N_c/V$, where $V$ is the volume accessible to the $N_c$ counterions~\cite{DaGu02}. In Fig.~\ref{fig8} we show the curves of $\phi_{osm}$ versus $\phi_{frac}$ for the same parameters as in Fig.~\ref{fig6} obtained with the present theory. For longer chains we can observe a minimum in the curve. Also, increasing the length of grafted chains we obtain a smaller osmotic coefficient which is agreement with experimental measurements~\cite{DaGu02}.

\section{Conclusions}

In this work we have performed MD simulations of a spherical polyelectrolyte brush in a salt free solution. The dielectric discontinuity in the grafted nanoparticle surface is taken into account. We observe that for monovalent counterions at room temperature the grafted nanoparticle polarization is not mandatory to describe the ionic structure around the brush. Also, the effective polyelectrolyte brush radius is not very affected for the studied parameters apart from the cases with trivalent conterions and longer chains, which differences can achieve $\approx 9\%$. Furthermore, in these cases, the concentration profiles of counterions and monomers is considerably different comparing both approximations. We also present a mean-field Poisson-Boltzmann theory for low electrostatic coupling regime. This method allows us to obtain quantitatively the asymptotic counterionic concentration, leading us to calculate the osmotic coefficients of PEBs suspensions. The effective charges of brushes are going to be studied in a future work.

\section{Acknowledgments}
This work was partially supported by the CNPq, CAPES and Alexander von Humboldt Foundation.

\bibliography{ref.bib}

\end{document}